\documentclass[twocolumn,showpacs,showkeys,superscriptaddress,preprintnumbers,amsmath,amssymb,prl]{revtex4}
\usepackage{graphicx}% Include figure files
\usepackage{dcolumn}% Align table columns on decimal point
\usepackage{bm}% bold math
\usepackage{amsmath}
\begin{document}

\title{Reynolds Pressure and Relaxation in a Sheared Granular System}

\author{Jie Ren}

\author{Joshua A. Dijksman}

\author{Robert P. Behringer}

\affiliation{Department of Physics \& Center for Non-linear and Complex Systems, Duke University, Science Drive, Durham NC 27708-0305, USA}

\date{\today}

\begin{abstract}
We describe experiments that probe the evolution of shear jammed
states, occurring for packing fractions $\phi_S \leq \phi \leq
\phi_J$, for frictional granular disks, where above $\phi_J$ there are
no stress-free static states.  We use a novel shear apparatus that
avoids the formation of inhomogeneities known as shear bands.  This
fixed $\phi$ system exhibits coupling between the shear strain,
$\gamma$, and the pressure, $P$, which we characterize by the
`Reynolds pressure', and a `Reynolds coefficient', $R(\phi) =
(\partial ^2 P/\partial \gamma ^2)/2$. $R$ depends only on $\phi$, and
diverges as $R \sim (\phi_c - \phi)^{\alpha}$, where $\phi_c \simeq
\phi_J$, and $\alpha \simeq -3.3$.
% $R$ shows a strong increase with increasing $\phi$ for $\phi \geq \phi_S$.
Under cyclic shear, this system evolves logarithmically slowly towards
limit cycle dynamics, which we characterize in terms of pressure
relaxation at cycle $n$: $\Delta P \simeq -\beta \ln(n/n_0)$.  $\beta$
depends only on the shear cycle amplitude, suggesting an activated
process where $\beta$ plays a temperature-like role.

\end{abstract}

\keywords{Granular materials, jamming, shear jamming}
\pacs{83.80.Fg, 62.20.D-, 83.85.Vb}

\maketitle

Much recent work has focused on the mechanical behavior of disordered
solids, including granular materials, colloids, foams and molecular
glass formers. These systems are well known for their glassy flow
behavior and surprising rigidity. Notably, Bi et
al.~\cite{zhang10,bi11} recently showed that in frictional systems,
e.g. most common granular materials, shear strain, $\gamma$, can
`shear jam’~\cite{bi11} a loose, low density packing of particles,
enabling it to support a shear stress. The nature of these shear
jammed states, particularly how they form and evolve, is an unsolved
problem with obvious relevance, whose understanding is the goal of the
current paper.

To set the context, we note that Bi et al.\cite{bi11} showed that
there is a lowest packing fraction $\phi_J$, such that below (above)
this density, there are (no) zero-stress states. Application of shear
to a zero-stress state in $\phi_S \leq \phi \leq \phi_J$ leads to
highly anisotropic contact and force networks, and to non-zero shear
stress, $\tau$, and pressure, $P$. Here, $\tau = (\sigma_1 -
\sigma_2)/2$ and $P = (\sigma_1 + \sigma_2)/2$, where the $\sigma_i$
are the principal stresses of the 2D stress tensor, $\hat{\sigma}$.
Starting from zero stress, the system traverses a fragile regime, and
with additional shear strain, the system arrives at a fully jammed
state where the force/contact networks percolate in all directions.
These shear jammed states may occur naturally in many granular
systems, such as geophysical flows, sand and suspensions. Improved
understanding of shear jammed states is thus crucial for both a better
understanding of the concept of jamming for (frictional) materials,
and to shed light on the complex rheology of dense granular
media~\cite{pouliquen08}.

At the heart of shear jamming are classic studies by Reynolds, who
showed that under fixed pressure, granular systems can dilate in
response to shear~\cite{reynolds}. Despite its relevance, a
quantitative understanding of this effect has remained elusive over
the last century. This is partly due to a complication in the study of
sheared frictional materials: Shear typically induces the formation of
dilated localized shear bands, where most of the shear strain is
confined.  System-wide measures may tend to reflect the band
properties rather than the whole system, making it difficult to
interpret experiments.  

To understand the important physics underlying shear jamming, it is
crucial to have an experimental approach that avoids shear banding.
In this Letter, we describe such an approach that, for the first time
to our knowledge, avoids shear banding.  Measurements using this
method provide the first characterizations of, and key insights into,
the mechanical response and dynamics of shear jammed frictional
packings.  In these fixed volume experiments, the response to shear is
manifested as a nonlinearly growing pressure with shear strain, which
is related to Reynolds' dilatancy. Associated with this pressure
effect are structural rearrangements that lead to a surprising
Arrhenius-like stress relaxation dynamics in periodically sheared disk
packings.

{\em Key Findings} In these experiments, we shear a disordered disk
packing (2D) at fixed density. In such a system, dilatancy cannot
occur, but a related phenomena occurs: the stresses $\hat{\sigma}$
respond to the shear strain. We find that $P$ increases roughly as
$\gamma^2$, which we describe by a ``Reynolds coefficient", $R =
(\partial ^2 P/\partial \gamma ^2 _{|\phi})/2$.  We find that $R$
depends only on $\phi$, and it provides a simple parametrization of
the coupling between $P$ and $\gamma$.  $R$ seems to diverge as $\phi$
approaches $\phi_c \simeq \phi_J$, thus identifying a special role for
$\phi_J$ for the shear jamming states.

An additional key observation from this work is that for $\phi_S \leq
\phi \leq \phi_J $ the stress response to cyclic shear strain shows
slow relaxational dynamics to a limit cycle, that depends on
driving. The deviation from a limit cycle, measured by pressure, shows
a logarithmic decay over time/cycle number.  The data for stress
relaxation exhibit a totally unexpected scaling form, as
developed below.

\begin{figure}[tbp]
\includegraphics[width=8.5cm]{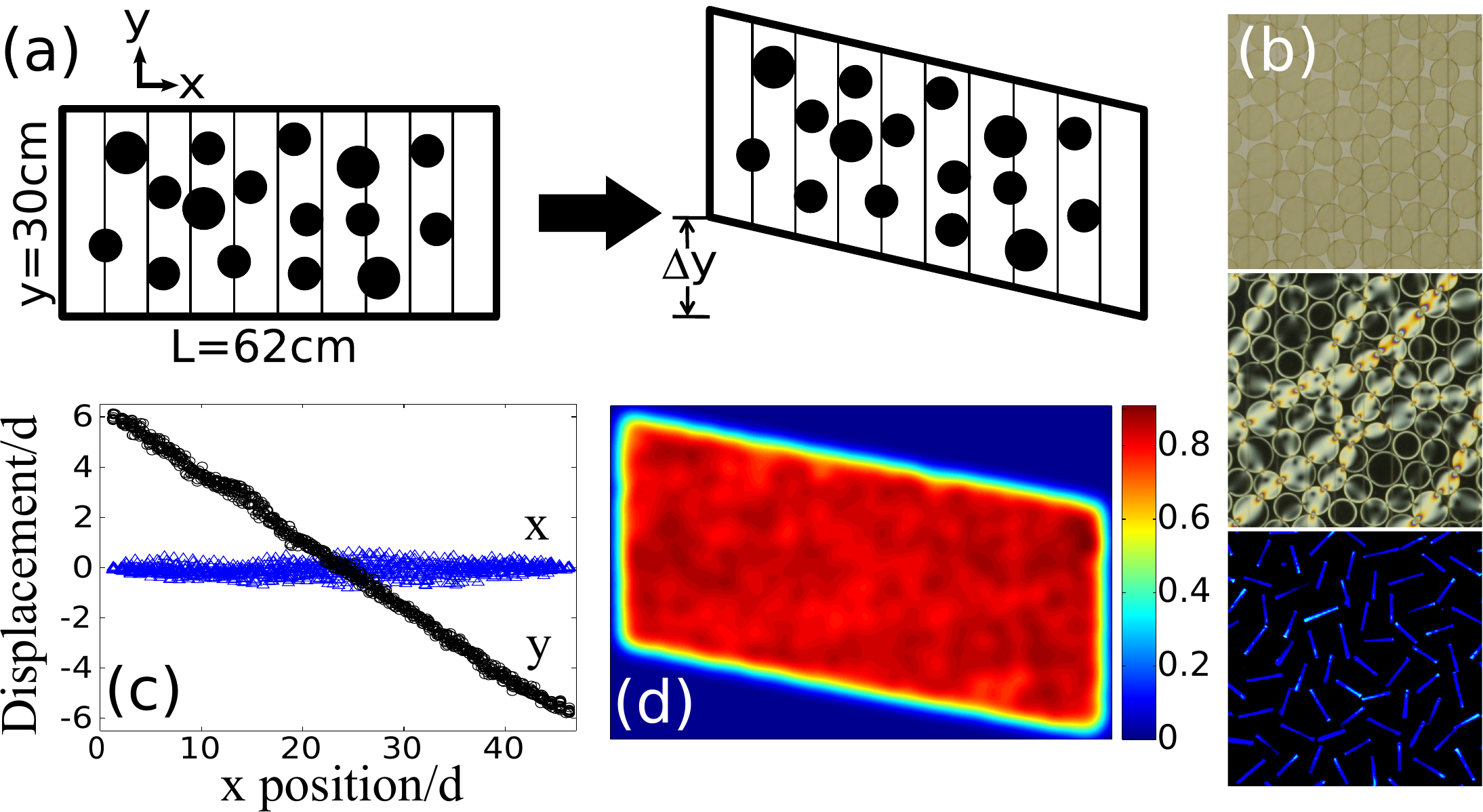}
\caption{\label{fig:setup} (Color online) (a) Setup schematics (b) The
  three close-up images that the camera captures at each step:
  particle positions (upper), force response under polariscope
  (middle), and particle orientation images under UV light
  (lower). (c) The $x-$ and $y$-displacement of particles vs. their
  horizontal positions in the system, and (d) the coarse
  grained~\cite{Zhang_CG,Clark_CG} density profile after 27\% linear
  shear.}
\end{figure}

\emph{Experimental Setup} Key to these experiments is a novel
apparatus that provides (simple) shear throughout the system, in
contrast to wall-driven shear.  The base of the apparatus consists of
narrow, parallel, horizontal, and transparent slats. Shear is applied
by deforming the slats and boundary uniformly in the `y' direction,
keeping the `x' dimension fixed at $L$, to provide uniform simple
shear strain $\gamma=\Delta y/L$ at constant packing fraction $\phi$
(Fig.~\ref{fig:setup}a). On the slats rest $\sim 1000$ bi-disperse
photo-elastic particles (Vishay PSM-4) of diameters $12.7mm$ and
$15.9mm$; the slat width is of the order of the particle size. The
relative numbers of large to small particles is set to $1:3.3$, in
order to prevent crystallization. Before each experiment, we
  prepare a stress-free packing by rearranging the particles (gently
  tapping or pushing particles) until no visual photoelastic response
  is visible. This bottom-assisted shear induces a linear shear
profile, suppressing shear bands and the usual inhomogeneities. It is
reminiscent of the SLLOD and related algorithms~\cite{SLLOD} for
enforcing uniform shear in MD simulations. It bears some resemblance
to 3D experiments by Mueggenburg~\cite{Mueggenburg}, but with a key
difference: in the Mueggenburg experiments, a slat geometry was used,
but the slot motion was not coordinated, and sustained uniform shear
did not occur.  We note that a small background pressure of $\sim
  0.5$~N/m is detected, even in the absence of shear.  This is due in
  roughly equal amounts to small experimental errors in force
  determinations, our ability to completely relax all inter-particle
  forces, and weak friction between the particles and the slats.

The experiment is illuminated from below by circularly polarized
uniform white light, and from above by a less intense UV light.  A 22
Megapixel camera above the experiment records views with and without a
circular polarizer.  We apply quasi-static shear strain in small
steps.  After each step, we pause and record three views of the system
which respectively yield particle positions, photo-elastic responses,
and rotations. Without crossed polarizers, the edges of the particles
are visible (Fig.~\ref{fig:setup}b upper), and we use a circular
Hough-transform technique~\cite{Peng1} to determine particle centers
with an accuracy of $\sim 0.02 d$. With a circular polarizer in front
of the camera, we image the photo-elastic pattern of colored fringes
within each particle, which encode the contact forces acting on each
particle (Fig.~\ref{fig:setup}b middle). To determine the particle
orientations, each disk is marked diametrically with a line of
fluorescent dye, visible under UV light with the white light turned
off (Fig.~\ref{fig:setup}b lower).  Changes in the bar orientations
give particle rotations. The complete process of multiple strain
steps, followed by imaging after each step, is fully automated, and we
record up to 500 shear cycles per run.  We extract the local particle
stress by either a pattern-fitting approach\cite{majmudar07,zhang10},
yielding the complete contact network, particle forces, and stress
tensor (e.g. $P$ and $\tau$), or via $G^2$, the local squared
intensity gradient of the photo-elastic response, averaged on each
particle~\cite{Howell,Geng}. $G^2$ is a one-to-one function of $P$ on
the particle level, providing an efficient measure for $P$. For
small/large data sets, we use the former/latter approach.

\emph{Reynolds Effect} As noted, a striking aspect of applying shear
strain to a stress-free state for $\phi_S \leq \phi \leq \phi_J$, is
the generation of nonzero $P$ and $\tau$, as in the shear jamming
experiments of Bi et al.~\cite{bi11}.  In the present
  experiments, we go well beyond Bi et al. to probe the {\em
    evolution} of shear jammed states, first by forward-shearing the
  system, and then by shearing cyclically.  Regarding forward shear,
we prepared packings in a stress free initial state, for $0.691 \leq
\phi \leq 0.816$, where $\phi_J = 0.835 \pm 0.005$, and $\phi_S
\approx 0.75$. We then quasi-statically shear the system by 200~small
strain steps of 0.27\%, up to a total strain of $\gamma =
54\%$~\cite{jie-chaos}.These experiments show shear
  jamming\cite{bi11}, as expected, but unlike previous experiments,
particle tracking data (Fig.~\ref{fig:setup}c) show that the shear is
effectively linear and homogeneous across the entire system. Particle
displacements and rotations relative to the uniform shear background
are small. The locally coarse-grained density field~\cite{Zhang_CG,
  Clark_CG} (Fig.~\ref{fig:setup}d), shows no sign of a shear band or
permanent inhomogeneities.

\begin{figure}[tbp]
\includegraphics[width=8.5cm]{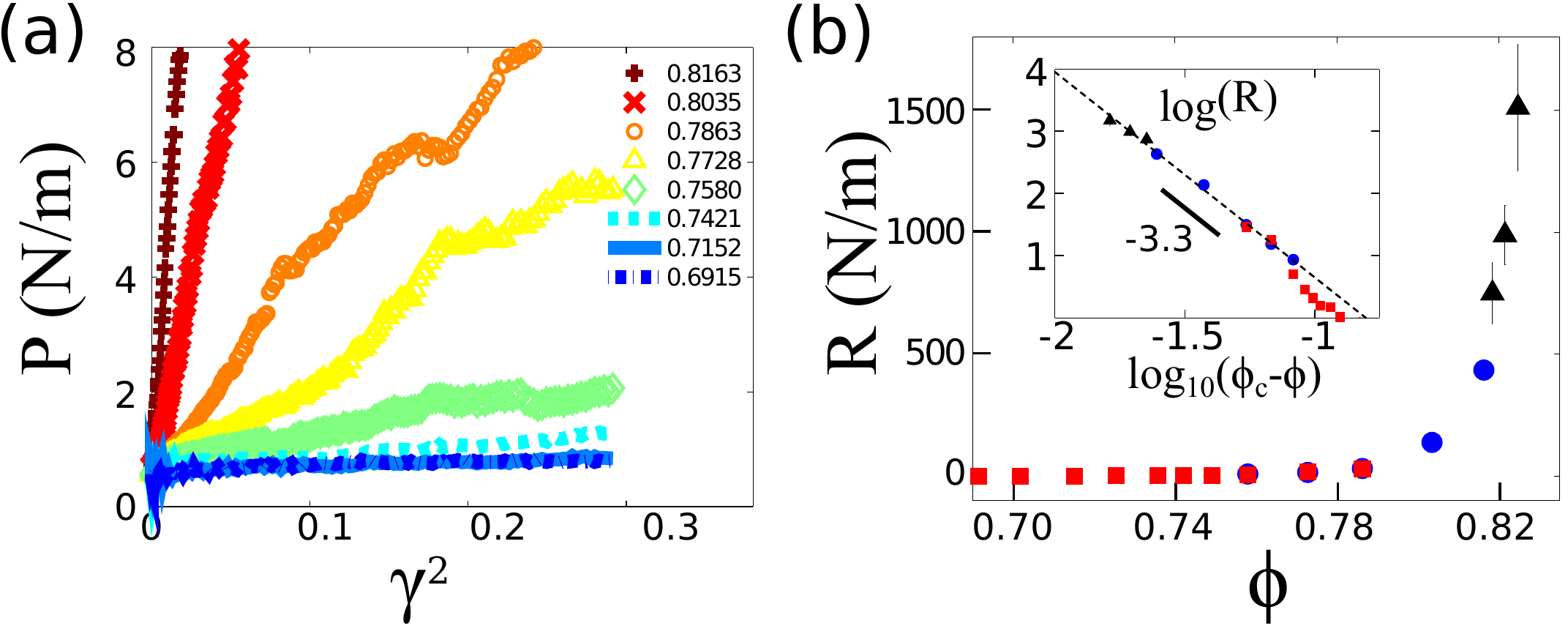}
\caption{\label{fig:dilate} (Color online) (a) Reynolds pressure
  $P(\gamma^2)$ observed in forward shear (see text) tests for $\phi =
  0.691 - 0.816$.  (b) Reynolds coefficient R extracted from linear
  fitting, obtained from up to 54$\%$ forward shear (red squares), up
  to 27\% forward shear (blue dots), and cyclic shear tests under
  limit cycle behavior (black triangles). The inset shows the same
  data on double logarithmic scales with $\phi_c = 0.841 \pm0.004$.
  The error bar is smaller than the size of the symbols unless marked.
  The dashed line shows a fit to a power law. A line corresponding to
  an exponent -3.3 is also shown for reference.}
\end{figure}

For the larger $\phi$'s considered here, we could not apply the full
$54\%$ strain because $P$ became so large that the layer was unstable
to out-of-plane buckling.  If buckling occurred, we terminated the
forward shear experiment.  The forward shear results,
Fig.~\ref{fig:dilate}a, indicate that the shear-induced `Reynolds
pressure' increases roughly as $\gamma^2$ with a density dependent
prefactor which we characterize by the `Reynolds coefficient',
\begin{equation}
R = (\partial ^2 P/\partial \gamma^2_{|\phi})/2.
\end{equation} 
For linear isotropic elastic materials, no coupling between shear
strain and pressure is expected.  But, as we apply shear, the system
becomes increasingly anisotropic, so a $P - \gamma$ coupling might be
possible, as expressed by, $\partial P/\partial \gamma$.  In our
system, this derivative grows roughly as $\gamma$, and linear
elasticity is not a particularly useful concept.  $R$ grows strongly
with $\phi$, and shows an apparent, but unexpected divergence at
$\phi = \phi_c \simeq \phi_J$.  Fig.~\ref{fig:dilate}b and inset, show
a log-log plot of $R$ vs. $\Delta \phi = \phi_c-\phi$.  A power-law
fit to $R = A(\phi_c -\phi)^{\alpha}$, yields $\alpha = -3.3 \pm 0.1$
and $\phi_c = 0.841\pm 0.004$.  By contrast, $\phi_c$ lies in the
range $0.83 \leq \phi_j \leq 0.84$, so here, $\phi_c$ is not
distinguishable from $\phi_J$, which is also comparable to $\phi_J$
for systems of frictionless 2D particles. For $ \phi \leq 0.75$, the
system is very loose, and it does not form a percolating contact
network, even after $54\%$ strain.  $R(\phi)$ behavior in this case
 is affected by small experimental `noise' effects, discussed
  above, and deviates from the power-law behavior
(Fig.~\ref{fig:dilate}b(inset)).  We identify $\phi_S \simeq 0.75$,
the lower limit in this system for shear jamming.

\emph{Limit Cycles} To characterize the
evolution/reproducibility/relaxation of the stresses, we carried out
multiple shear cycles. This also allowed us to determine $R$ for
$\phi$ closer to $\phi_J$, where shear strains are limited due to
buckling; we obtain good statistics by many smaller-amplitude strain
cycles.  The oscillatory shear experiments were started from initially
stress-free states for $\phi$'s in the shear jamming regime, $\phi_S
\leq \phi \leq \phi_J$.  In a cycle, we sheared by strain steps of
$0.45\%$ up to $\gamma_{max}$ in the `forward direction', followed by
a shear strain decrease ($-0.45\%$ per step) to a smaller strain,
$\gamma_{min}$. For symmetric shear cycles: $\gamma_{min} = -
\gamma_{max}$, and asymmetric shear cycles: $\gamma_{min} \ne -
\gamma_{max}$.

\begin{figure}[tbp]
\includegraphics[width=8.5cm]{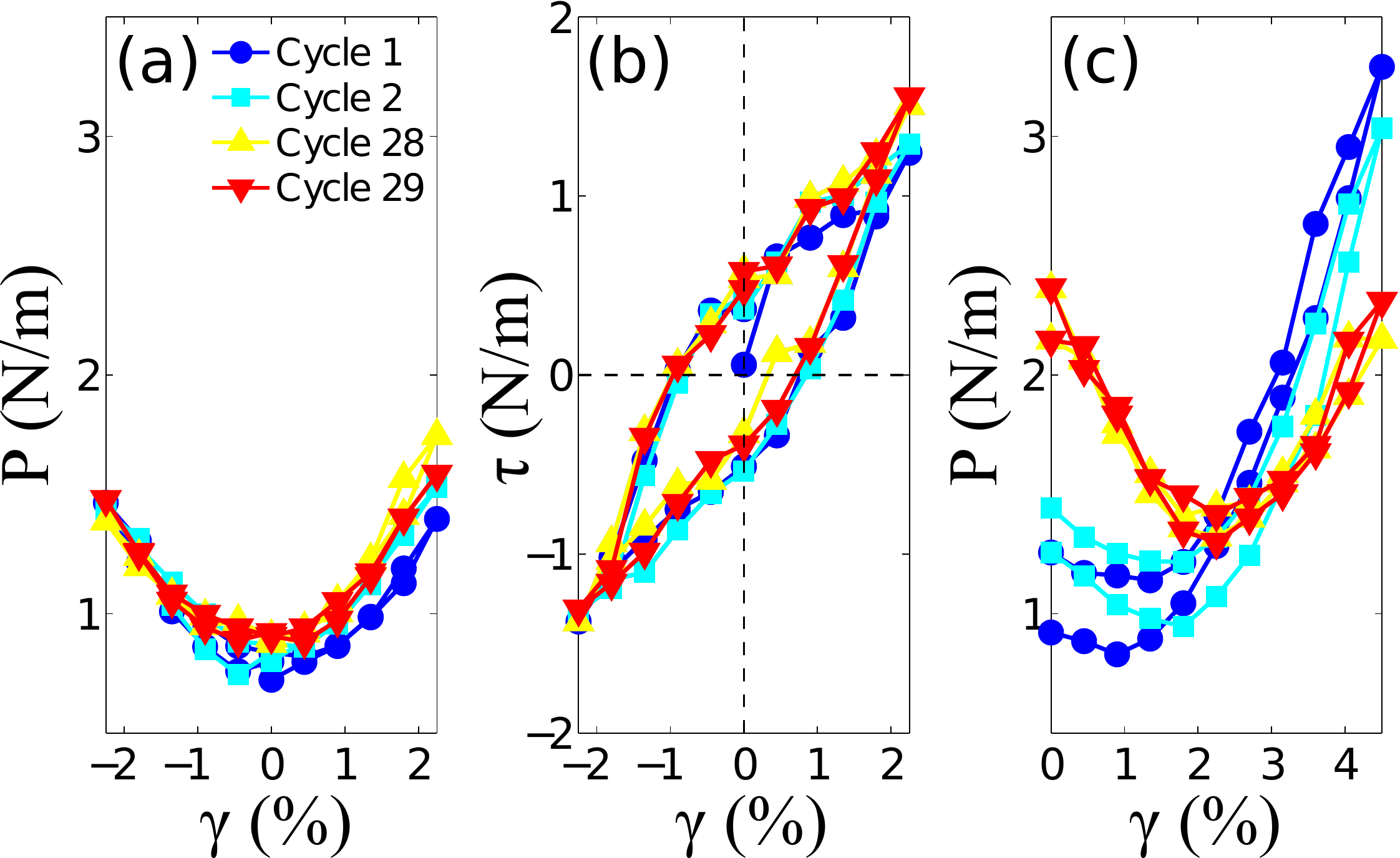}
\caption{\label{fig:multistep} (Color online) (a) $P$ vs.~$\gamma$ for
  a symmetric cyclic shear run with $\phi = 0.825$, which started from
  $\gamma=0$, and sheared between $\gamma_{max} = 2.25\%$ and
  $\gamma_{min} = -2.25\%$.  Only cycles 1, 2, 28 and 29 are shown in
  the plot.  (b) $\tau$ vs. $\gamma$ for the same run and the same
  shear cycles. (c)$P$ vs.~$\gamma$ at cycles 1, 2, 28, 29 for a
  non-symmetric cyclic shear run ($\gamma_{max} = 4.5\%$,
  $\gamma_{min} = 0$) with the same density.}
\end{figure}

For symmetric cycles, $P$ was symmetric about $\gamma = 0$,
approximately quadratic in $\gamma$, and virtually reproducible over
many cycles, as shown in Fig.~\ref{fig:multistep}a.  However,
  details of the network were generally not reproducible from cycle to
cycle. The Reynolds
coefficient $R(\phi)$ followed the same trend as in the forward shear
tests (Fig.~\ref{fig:dilate}b), further confirming the Reynolds
effect.  After transients, the shear stress $\tau$ also followed a
reproducible path over cycles, but unlike $P$, $\tau$ was strongly
hysteretic, with non-zero values at $\gamma = 0$. There were
$\gamma$'s for which $\tau = 0$ but $P \ne 0$, for example, in
Fig.~\ref{fig:multistep}a and b, at $\gamma \approx 1\%$. However, in
such cases, $\tau$ coarse grained at smaller scales than the system
size was \emph{locally} non-zero, even though the global $\tau$ was
$0$ (e.g. because of spatial variations of the principal stress
orientations).  Due to length limitations, we consider only the
dynamics exhibited by $P$, and we will present the full stress
dynamics elsewhere.

The evolution of $P(\gamma)$ for asymmetric shear cycles differed from
the symmetric case.  Here, $P(\gamma)$ was initially asymmetric, but
evolved towards a symmetric shape centered around the mean strain,
$\bar{\gamma}$, after many cycles. Thus, the long term $P - \gamma$
dynamics was a limit cycle.  The system relaxed quickly (slowly) to
the limit cycle if sheared symmetrically
(asymmetrically). Fig.~\ref{fig:multistep}c shows an example of slow
evolution, where a limit cycle was reached after about 28 cycles.  In
this case $P(\gamma)$ evolved to a symmetric shape, similar to the
forward shear experiment, except for a shift; i.e., the system did not
reach a completely stress-free state at the mid-point of
strain. However, a long term limit cycle was still reached with the
same Reynolds coefficient for the given density, $\phi = 0.825$.

\begin{figure}[tbp]
\includegraphics[width=8.5cm]{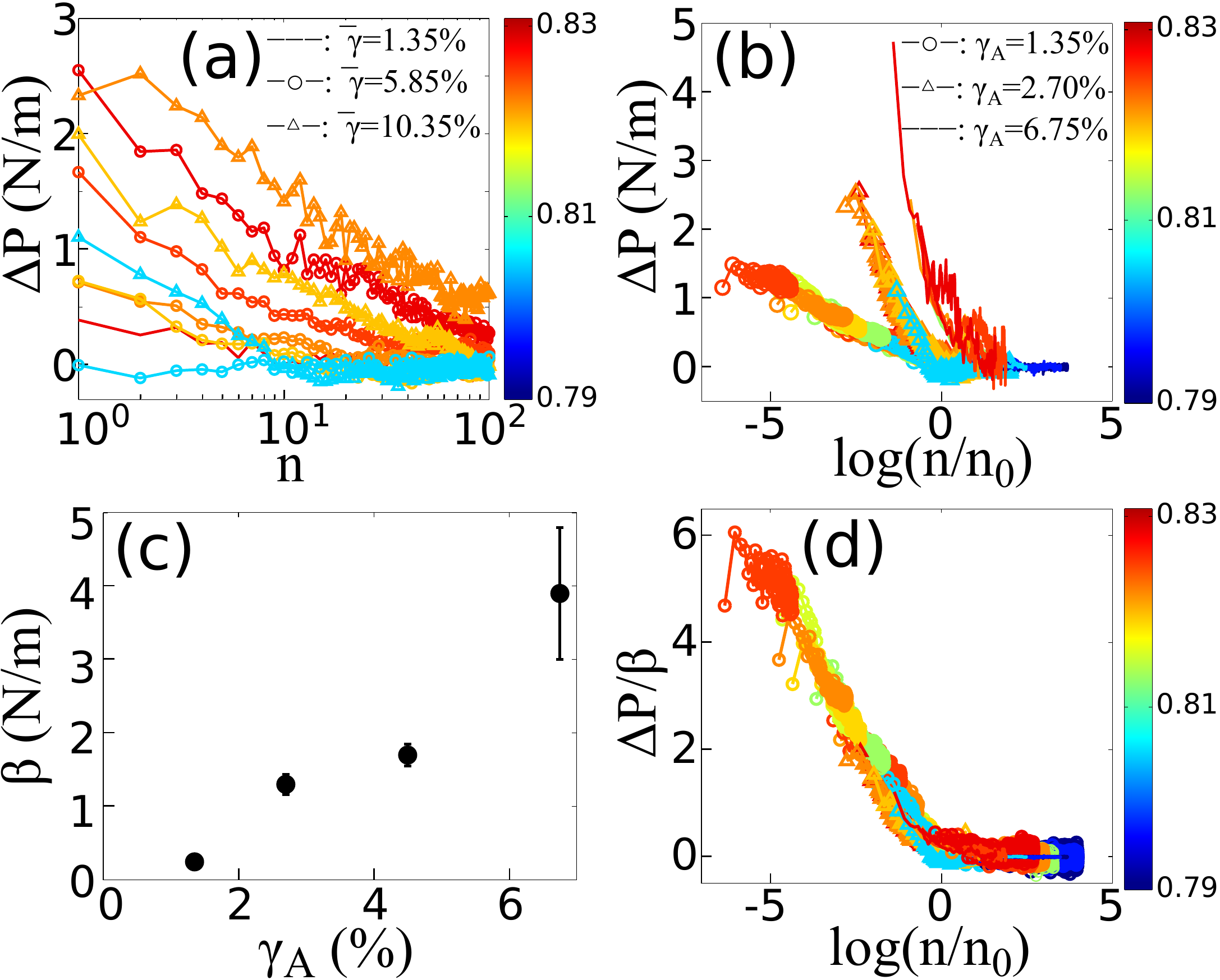}
\caption{\label{fig:relax} (Color online) (a) $\Delta P$ vs.~number of
  shear cycles, for various runs with $\gamma_A = 2.7\%$, and
  $\bar{\gamma} = 1.35\%$(solid line), $5.85\%$(open circles), and
  $10.35\%$(open triangles). Different color indicates different
  $\phi$.  (b) $\Delta P$ vs.~$log(n/n_0)$ for 3 different
  $\gamma_A$'s: 1.35\%(open circles), 2.70\%(open triangles), and
  6.75\% (solid lines). Data comes from runs with various $\phi$
  (shown by color) and various $\bar{\gamma}$(not shown).  (c) The
  decay factor, $\beta$, shows a strong increase with $\gamma_A$.  (d)
  The universal decay curve, $\Delta P/\beta(\gamma_A)$
  vs.~$log(n/n_0)$, including all data from panel (b).}
\end{figure}

\emph{Slow Relaxation} For asymmetric strain cycles, $\Delta P(n) =
P(\gamma_{max}) - P(\gamma_{min})$ was initially nonzero, but it
decreased and ultimately vanished, within fluctuations, for $n = n_0$.
When the limit cycle was reached, $P$ was symmetric about $\bar{\gamma}
= (\gamma_{max} + \gamma_{min})/2$.  The slow relaxation of $\Delta P$
for asymmetric shear shows striking and novel scaling behavior, which we
characterize in terms of $\phi$, $\bar{\gamma}$ and the shear
amplitude $\gamma_A$.  Experiments to characterize this relaxation
spanned $\phi$'s from above $\phi_S$ to just below isotropic jamming
$\phi_J$: $0.780 \leq \phi \leq 0.828$, strain amplitudes of $\gamma_A
= 6.75, 4.5, 3, 1.5\%$ and a range of starting strains $0 \leq
\bar{\gamma} \leq 21.35\%$. Experiments were 100-500~cycles long; for
convenience we measured $G^2$ only at $\gamma_{max}$, $\gamma_{min}$,
and then converted $G^2$ to $\Delta P$ using a
calibration. Fig.~\ref{fig:relax}a shows $\Delta P$ for a particular
$\gamma_A$.

For $\phi$ in the shear jamming region, $\Delta P(n)$ decayed
logarithmically slowly towards $0$:
\begin{equation}
\Delta P(n) \simeq -\beta \log(n/n_0),
\label{eq:log-scaling}
\end{equation}
implying a natural `time scale' for relaxation, $n_0$, that we
obtained through least squares fits of the logarithmic part of the
relaxation.  All the relaxation data, for a given $\gamma_A$, collapse
onto a single curve when expressed in terms of $n/n_0$
(Fig.~\ref{fig:relax}b), regardless of $\phi$ and $\bar{\gamma}$.  The
factor $\beta(\gamma_A)$ differs for each $\gamma_A$
(Fig.~\ref{fig:relax}c), but $\Delta P/\beta$ is a universal function
of $n/n_0$, as in Fig.~\ref{fig:relax}d, which shows all
$\sim$170~datasets.  We emphasize the remarkable role that
$\beta(\gamma_A)$ plays, and the fact that it is independent of
$\phi$.

We then consider what determines $n_0$. Eq.~\ref{eq:log-scaling}
implies: $n_0 = n \cdot \exp (\Delta P(n)/\beta(\gamma_A))$.  Initially, at $n = 1$, 
$\Delta P = \Delta P_0$. According to the approximately quadratic relation 
between $P$ and $\gamma$, $\Delta P_0$ is given by:
$\Delta P_0 = R(\phi)(\gamma_{max}^2 - \gamma_{min}^2)/2 =
R(\phi)\bar{\gamma}\gamma_A$. Therefore,
\begin{equation}
n_0 = \exp(R(\phi)\bar{\gamma} \frac{\gamma_A}{\beta(\gamma_A)}).
\label{eq:timescale}
\end{equation}

\noindent
Eq.~\ref{eq:log-scaling} also implies an evolution $d \Delta P / dn =
-\beta n_0^{-1} \exp(\Delta P /\beta)$ or, with a cutoff, $d \Delta P
/ dn = -\beta n_0^{-1} [\exp(\Delta P /\beta) - 1]$, which produces
the logarithmic form of Eq.~\ref{eq:log-scaling} for small $n$, with
saturation at $n = n_0$.  This suggests an activated process, perhaps
involving a generalized ensemble, such as the stress ensemble, as
discussed by several
authors\cite{pouliquen-activated1,behringer08,stress-ensemble,sollich}.

To summarize: for frictional granular systems in/near the shear
jamming regime, $\phi_S \leq \phi \leq \phi_J$, we generated sheared
states without shear bands, even with large strains or over many
cycles of shear, making it possible to experimentally probe the
constitutive relations of granular materials.  These experiments show
two key and highly novel results: 1) We find a novel Reynolds effect
for fixed $\phi$ that is approximately quadratic in $\gamma$ using $R
= (\partial ^2 P/\partial \gamma^2_{|\phi})/2$. We note that the
specific form for $R(\gamma)$ may well depend on the particle
interaction force; a more general form might be $P = R
\gamma^{\delta}$, where for our experiments, $\delta \simeq 2$.  2) We
find that under cyclic shear, frictional granular systems evolve
logarithmically slowly, as one might expect for an activated process,
towards a state where the pressure is symmetric, {\em modulo}
fluctuations, about the mid-point of strain.  The pressure at the
symmetry point may not be zero.  This slow evolution is characterized
by highly novel scaling behavior, such that there is good collapse of
all data.

These results point towards several interesting directions.  First, it
is reasonable to search for a description of these states in terms of
an ensemble picture, such as the stress ensemble, given the activated
process character of the slow relaxation.  Such a theory would need to
explain some of the striking scaling properties observed here.  In
addition, we have not considered the properties of the shear stress
under cyclic shearing, nor have we considered the particle dynamics
of details of the force/contact networks.  We will present these
results elsewhere.

We thank Jie Zhang for sharing code to perform rotational particle
tracking. IGUS generously supplied us with a free linear stage under
the Young Engineers Support program. Discussions with Dapeng Bi,
Bulbul Chakraborty, Martin van Hecke, Stefan Luding and Corey O'Hern
are gratefully acknowledged. Work supported by NSF grants DMR-0906908,
DMR-1206351, ARO grant W911NF-11-1-0110, and NSF grant DMS0835742.

%\bibliography{granular}

\end{document}